**Mapping the long-term trajectories of political violence in Africa**

Steven M. Radil [a]*, Nick Dorward [b], Olivier Walther [a], Levi John Wolf [c]

March 6, 2026

[a] Department of Geography, University of Florida, USA.
[b] School of Geography and Environmental Science, University of Southampton, UK.
[c] School of Geographical Sciences, University of Bristol, UK.

*Corresponding author: steven.radil@ufl.edu

**Abstract:** Existing models of political violence often emphasize discrete transitions—when conflicts emerge, escalate, or subside—without considering the longer trajectories of violence that accumulate across time and space. This paper introduces a spatially explicit longitudinal sequence analysis to address this gap. Using event-level data from the Armed Conflict Location and Event Dataset covering Africa from 1997 to 2024, we classify locations according to the intensity and spatial concentration of violence, tracing how these states evolve into distinct conflict trajectories. Applying optimal matching and clustering techniques, we identify six recurrent patterns ranging from short-lived, localized outbreaks to protracted high-intensity conflicts. We further assess how these trajectories align across neighboring areas, revealing evidence of spatial interdependence, particularly in border regions. By highlighting the temporal rhythms and geographic linkages of political violence, the study advances conflict research beyond isolated transitions and provides a framework for understanding the life cycles of violence.

**Acknowledgements:** Funding for this work was provided by the Economic and Social Research Council (grant number ES/Y007840/1) and the Organisation for Economic Co-operation and Development (grant number AWD09867).

## 1. Introduction

Understanding how political violence unfolds across space and time remains a central challenge in conflict research. While decades of work have examined where and when violence occurs, much of this scholarship has focused on discrete transitions or moments when violence appears, intensifies, or disappears in specific locations. These approaches have generated valuable insights into conflict onset, diffusion, and escalation, but they often neglect the longer trajectories of violence: how sequences of violent and non-violent episodes accumulate over time to shape the life cycle of a conflict within a place. Without this broader temporal framing, it is difficult to know whether an observed shift represents a short-lived irregularity or a deeper structural transformation in a conflict's evolution.

This gap is especially relevant in the African context. Over the past two decades, political violence has become more geographically dispersed, complex, and persistent, involving insurgencies, communal clashes, and transnational armed groups. Research has documented how conflict diffuses across borders and how particular geographies, such as borderlands and road networks, can facilitate the spread of violence. Yet we know less about whether an outbreak of violence in a place follows regular temporal and spatial trajectories and whether the trajectory of one locality is connected to that of its neighbors. Addressing these questions is critical for distinguishing fleeting episodes of instability from enduring zones of conflict, and for designing interventions that respond to the actual space-time rhythms of violence.

This study introduces a spatially explicit longitudinal sequence analysis to conflict research. Originally developed in bioinformatics, sequence analysis is increasingly applied to social and geographical phenomena, but it remains underused in the study of political violence. By classifying the evolving states of conflict intensity and spatial concentration within each location, and by comparing these sequences across neighboring areas, we can identify common patterns in the emergence, persistence, and resolution of violence.

Two guiding questions shape the analysis. First, do conflicts follow distinct spatial and temporal trajectories within places, from their emergence to their resolution? Second, are the trajectories of conflict in one location associated with those of neighboring areas, suggesting processes of spatial interdependence or diffusion? Answering these questions allows us to move beyond isolated event transitions and toward a more integrated understanding of conflict life cycles.

The paper proceeds as follows. The next section reviews the existing literature on conflict trajectories and diffusion, highlighting both achievements and important gaps that inform our effort. We then describe the methodological framework of spatially explicit sequence analysis and its application to conflict event data in Africa from 1997 to 2024. The results section presents a typology of conflict trajectories and examines their spatial distribution. The discussion reflects on what these findings imply for theories of conflict diffusion and persistence, before the paper concludes with implications for future research and policy.

## 2. Literature review

### *2.1. Trajectories of armed conflict*

The growing availability of georeferenced conflict datasets has reshaped the study of civil wars and political violence. Whereas earlier generations of scholarship often relied on country-level or annual data, fine-grained event datasets allow scholars to examine how violence unfolds across both space and time (Levin et al., 2018; Miller et al., 2022). Accordingly, geographic approaches to conflict dynamics have figured prominently in the literature, yielding insights into the shifting geographies of civil wars (Schutte and Weidmann, 2011), the changing strategic concerns of conflict actors (O'Loughlin and Witmer, 2012), the efficacy of counter insurgency strategies vis-à-vis the spread and escalation of violence (Toft and Zhukov, 2015), and the changing geographies of conflict over time (Walther et al., 2023).

This shift has also produced a large body of subnational research linking the onset, frequency, and intensity of political violence to actor dynamics (Giraudy et al., 2019) and how geographic factors, such as ethnic settlement patterns, economic inequalities, or natural resource endowments condition the likelihood of conflict onset and duration (Buhaug et al., 2009; Raleigh & Hegre, 2009; Raleigh & Linke, 2018).

Parallel work in terrorism studies similarly emphasizes geographic determinants of violence. Forest cover, distance to borders, and population density have all been employed to explain and predict episodes of violence (Braithwaite & Li, 2007; Marineau et al., 2020). This predictive orientation has culminated in forecasting efforts such as the Violence Early Warning System (ViEWS), which uses lagged event data to predict risks of conflict across Africa (Hegre et al., 2019). The overarching goal of this research agenda has been to identify the correlates of violence and to develop models capable of anticipating its future occurrence.

While such approaches have generated valuable findings, they tend to conceptualize violence as a series of discrete transitions, identifying moments when a location shifts from peace to conflict, or from low to high intensity. Regression-based frameworks, in particular, are oriented toward correlating the presence or absence of violent events with a set of assumed exogenous covariates. Although effective for identifying correlations and improving prediction, these methods make it difficult to determine whether observed transitions are isolated irregularities or parts of broader, longer trajectories of violence.

This limitation has become increasingly apparent in research on the geography of African conflicts. Studies of North and West Africa, for example, show that violence often begins and ends in dispersed forms but tends to cluster spatially when it persists across years (OECD, 2020). Walther et al. (2023) formalized this observation through the Spatial Conflict Dynamics indicator, which captures how patterns of intensity and clustering evolve within localities. Such findings highlight that conflicts in Africa follow recognizable "life cycles" marked by recurring temporal and spatial stages (Walther et al., 2025). Yet the dominant modeling frameworks rarely capture these life cycles in their entirety, focusing instead on the likelihood of transitions at particular points in time. What remains missing is a systematic account of the *trajectories* of conflict—the sequences through which violence emerges, stabilizes, and eventually dissipates within a given place.

### *2.2. Sequences of armed conflict*

A smaller body of work explicitly treats conflict as a dynamic process unfolding across both space and time. Studies of diffusion have been especially influential in showing how violence spreads from one location to another, either through relocation (armed groups moving to establish bases in new areas) or expansion (the extension of territorial control) (O'Loughlin & Raleigh, 2008; Schutte & Weidmann, 2011; Bormann & Hammond, 2016; Metternich et al., 2017). These studies underscore that violent events are not independent but display strong spatial-temporal dependencies.

Border regions are critical spaces in this literature. They often function as recruitment zones, logistical corridors, and safe havens for armed groups seeking to evade state control (Arsenault & Bacon, 2015). Groups such as Al Qaeda in the Islamic Maghreb, Boko Haram, and the Lord's Resistance Army have long exploited porous borders to sustain operations and expand their reach (D'Amato, 2018; Dowd, 2017). Such findings highlight how geographic interdependencies shape the spread and escalation of violence. Yet they still conceptualize conflict primarily in terms of discrete state transitions (violent versus non-violent) observed across short intervals.

Two key limitations follow. First, approaches that aggregate transitions across locations or time struggle to differentiate between short-lived irregularities and deeper, more persistent shifts in conflict dynamics. For example, a single violent flare-up may appear indistinguishable from the early stages of a longer conflict cycle. Second, most studies treat each transition in isolation, thereby detaching it from the wider history of violence within a locality. As a result, the temporal ordering of events and how local trajectories connect to those of neighboring areas remain underexplored.

An alternative is to conceptualize conflicts as *sequences* of evolving states. Sequence analysis, originally developed in bioinformatics, provides tools for capturing ordered event histories and has been increasingly applied in social science. Scholars have used it to study labor market careers, neighborhood change, and migration pathways, demonstrating how trajectories emerge from the sequencing of states rather than isolated transitions (Delmelle, 2016; Vogiazides & Mondani, 2023; Losacker & Kuebart, 2024). Conflict research has only rarely adopted this approach (Stehle & Peuquet, 2015), but the logic is well suited to the problem since conflicts unfold as temporally ordered processes that can be compared, clustered, and typologized.

Bringing sequence analysis into conflict research allows scholars to trace how specific places move through different states of violence over time, distinguishing ephemeral outbreaks from enduring cycles. It also provides a framework for comparing local trajectories, identifying recurring patterns across diverse contexts. Finally, it makes it possible to analyze how the trajectories of one location align with or diverge from those of its neighbors, thereby integrating concerns with diffusion and interdependence into a temporal sequence perspective.

Building on this literature, this paper argues that understanding the trajectories of political violence requires methods that capture both temporal ordering and spatial association. Prior research has shown that violence spreads across space and recurs across time, but it remains unclear whether these processes coalesce into identifiable, recurrent patterns of conflict life cycles. To address this, we employ a spatially explicit longitudinal sequence analysis of political violence in Africa between 1997 and 2024. This approach enables us to ask two questions that existing frameworks leave unanswered:

1. Do conflicts follow distinct spatial and temporal trajectories within places, from their emergence to their resolution?
2. Are the trajectories of conflict in one location systematically associated with those of neighboring areas, revealing patterns of diffusion and interdependence?

By embedding conflict events within sequences rather than transitions, and by situating those sequences in their neighborhood context, this study provides both a methodological innovation and a theoretical contribution. Methodologically, it demonstrates how sequence analysis can be adapted to the study of armed conflict. Theoretically, it reframes conflict not as a set of independent transitions but as an evolving life cycle, shaped by both local conditions and spatial linkages across regions.

## 3. Data and methods

Our spatially explicit sequence analysis entails three main steps (Figure 1). First, we specify a state space and derive the empirical transitions grid cells take between these states. Second, we use optimal matching to measure the similarity in sequences across all pairs of grid cells. Then, we cluster cells according to their sequence similarity, providing a typology of cells' conflict experiences within Africa. From this solution, we ground our interpretations in maps of the cluster types, using join counts and other statistics to confirm this visual intuition.

Figure 1. Deriving distinct long-term trajectories of political violence

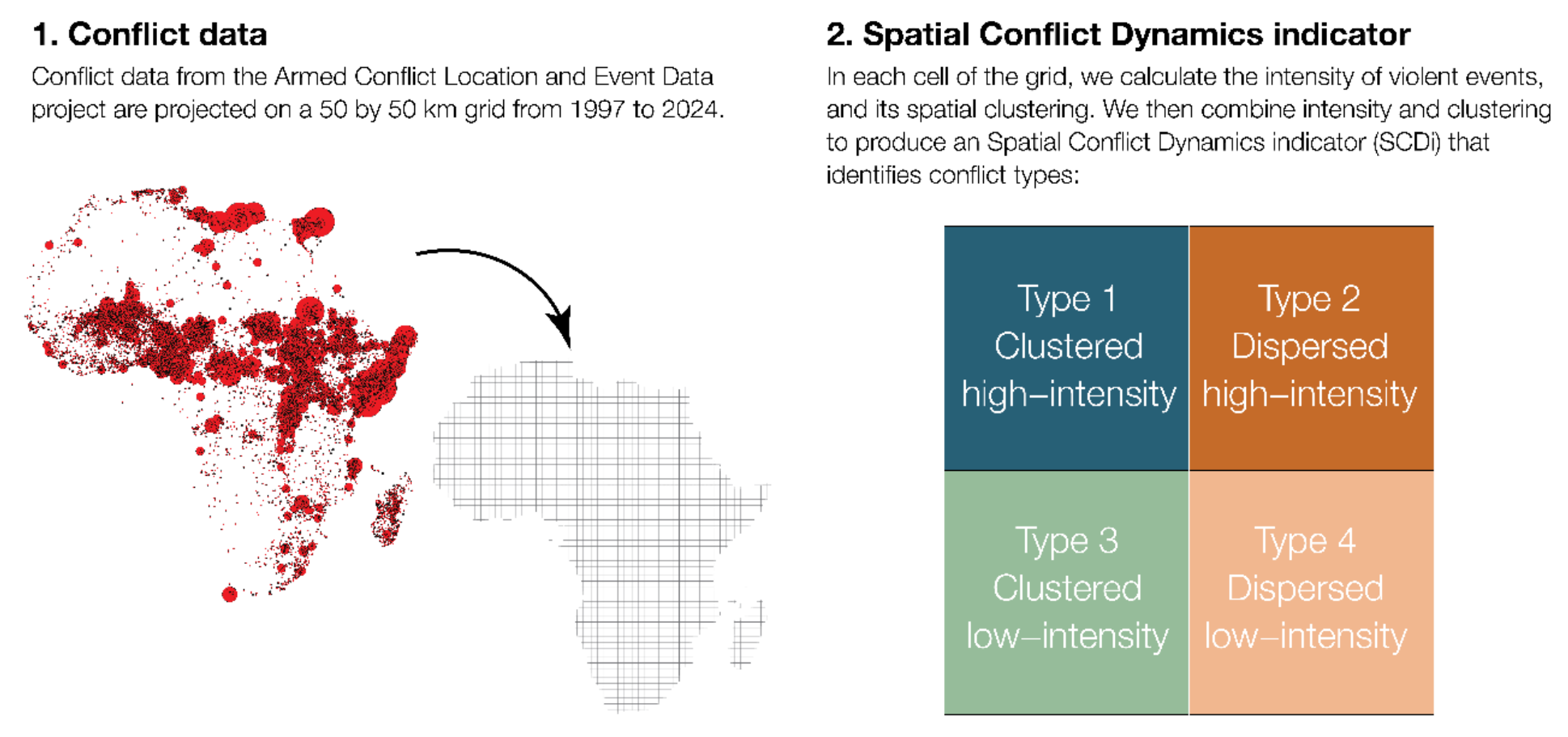

**3. Clustering analysis**

We then apply the Optimal Matching algorithm (OM) which uses the edit distances required to transform one sequence into another. Hierarchical Ward clustering then gathers cells with similar sequences together into six different cell types.

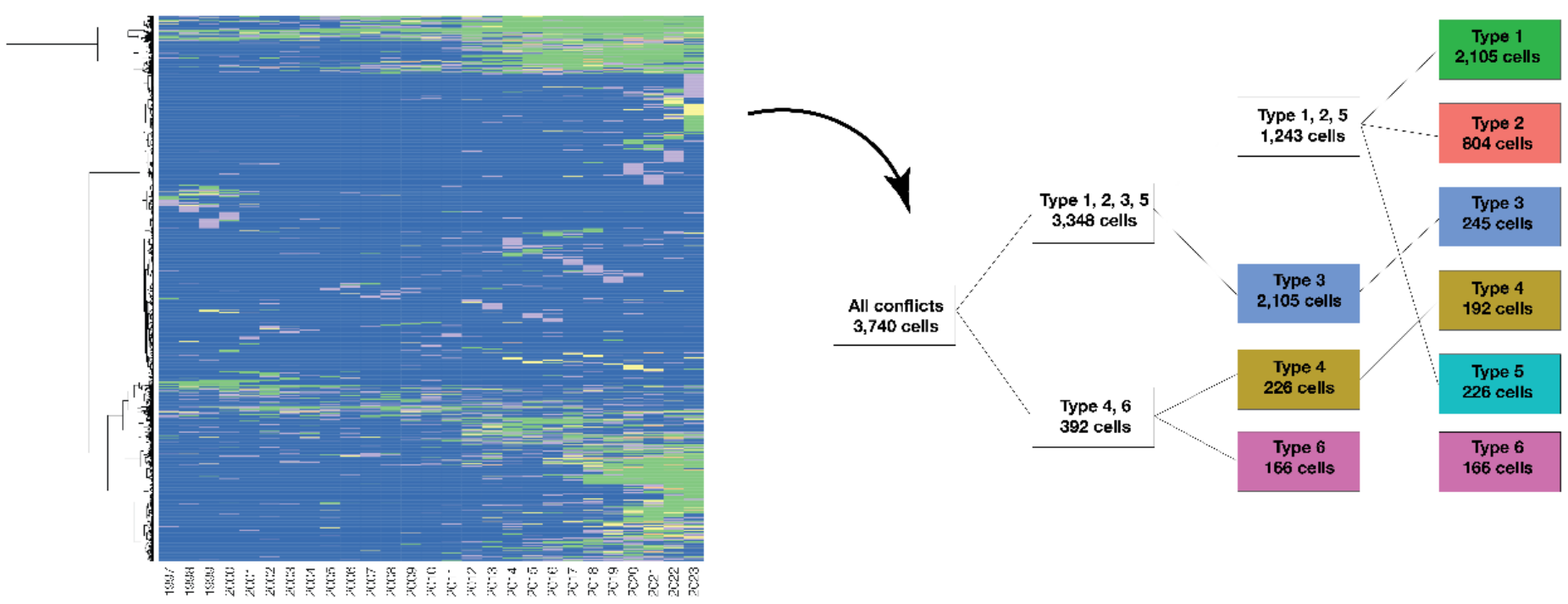

Source: authors.

The first decision in this analytical process is to specify how the set of states that grid cells can occupy. It is rare that clear exogenously determined states exist for many applications, and violent conflict is no different. In social science, state spaces are often synthesized from demographic data using unsupervised learning (Delmelle, 2016). In contrast, we turn to theory in order to define our state space, considering the location of individual violent events according to the intensity and concentration of the events themselves.

### *3.1. Conflict data: The Armed Conflict Location and Event Dataset*

For the event data itself, we draw upon data from the Armed Conflict Location and Event Dataset (ACLED), which provides georeferenced event-level data on a range of violent and non-violent events including riots, protests, violence against civilians, explosions and battles between state and non-state groups (Raleigh et al., 2010). These data conceptualize episodes of conflict as a series of discrete events representing contentious interactions between specific actors at a given location and time. ACLED codes events from local, regional, and national media; NGO reports; and, for African data, Africa-focused news reports and analysis. We use the full time series from 1997 to the last completed year, 2024 and subset events based upon the ACLED 'event type' variable. We keep the 'battles', 'explosions and remote violence', and 'violence against civilians' event types, excluding 'riots', 'protests', and 'strategic developments' which are not the focus of the study.

ACLED is one of a set of alternative event-level datasets, including the Uppsala Conflict Data Programme Georeferenced Event dataset (UCDP GED) (Sundberg and Melander, 2013) and the Social Conflict analysis Database (SCAD) (Salehyan et al., 2012). ALCED provides a reasonable compromise between breadth, reliability, and conceptual focus. First, ACLED includes a broader range of political violence than, for example, the GED which exclusively records instances of organized violence within formal armed conflicts. While some argue against this breadth of inclusion (Eck, 2012), ACLED is an important tool to understand the way places share experiences of political violence over time. Second, while SCAD includes a similarly broad range of contentious action, ACLED's coding of events is geographically more precise and less prone to reporting bias (Demarest and Langer, 2022).

### *3.2. Defining the state space using ACLED and the Spatial Conflict Dynamics indicator*

The Spatial Conflict Dynamics indicator (SCDi) is a general measurement strategy for representing and analyzing conflict dynamics in space and time (Walther et al., 2023). SCDi measures the intensity and concentration of conflict events across the study area to yield a five-class state space – locales may either be experiencing spatially 'clustered' or 'dispersed' violent events at 'high'- or 'low-intensity'.

- Clustered/high-intensity (CH) corresponds to areas with an above-average intensity and a clustered distribution of violent events, suggesting that violence is more frequent and happens near other events.
- Dispersed/high-intensity (DH) are areas with a higher-than-average intensity and a dispersed distribution of events, indicating that the frequency of violence is accelerating while the locations of violence are still quite distant from each other.

- Clustered/low-intensity (CL) can be found in areas where there are fewer violent activities and most of them take place near each other, indicating a decreasing conflict or one that is highly localized.
- Dispersed/low-intensity (DL) are areas where a lower-than-average intensity and a dispersed distribution of events are combined, suggesting that a conflict is lingering.
- Finally, some regions are coded No Conflict (NC) if no violence is observed in this locality.

This collection of states describes the way the spatial pattern of conflict events evolves spatially and temporally and is informed by the structure of the event pattern itself, rather than referring to violent events occurring in a particular collection of place types, like city centers or suburbs.

These types are indicative of potentially different stages in the overall lifecycle of a conflict (Walther et al., 2023; 2025). For example, when violence first emerges in a zone, the majority of cases are clustered and low-intensity (CL) and one third are clustered/high-intensity (CH). This indicates that violence is most likely to be concentrated spatially when it first emerges. However, once a conflict is established, it commonly persists over time in its clustered/high-intensity form. As conflicts start to end, they tend to move from CH and to DH before stopping altogether. Although violence has been observed to both initiate and end from all of the SCDi typologies, the dispersed categories (DH and DL) are most common either at the beginning or ending of a sequence of violence in a sub-zone. Further, dispersed conflicts are unlikely to persist compared with clustered conflicts and tend to dissipate quickly once they have emerged. This suggests that zones displaying these spatial typologies are either quite near the early stages of a conflict episode or the end. Finally, conflicts most commonly end by transitioning from CL to no conflict (NC) in the following year. This suggests that violence is often concentrated even just before it ends.

Taken together, the four spatial conflict categories reveal insights about the dynamics of the lifecycle of a typical conflict in Africa. These are general trends, however, and not all sub-zones, places or localities will exhibit the same life cycles between the SCDi categories. Nonetheless, a predominant pathway is suggested by the event data across the zones since the late 1990s. Emerging conflicts tend to result in clustering of either type, dispersed conflicts tend to quickly change, CH are more persistent, and violence most commonly ends from the CL forms.

### *3.3. Measuring sequence similarity using Optimal Matching*

From this state space, a *sequence* can be built and analyzed. A *sequence* is the collection of transitions a grid cell makes between states in a discrete time period. For instance, if a grid cell experiences CH conflict in 2002, CL conflict in 2003, and then NC in 2004, its sequence over this period is CH-CL-NC. Once the sequences are defined, we use the Optimal Matching (OM) algorithm to model the similarity between sequences using the number of edits it takes to transform sequences into one another (Kang et al., 2020).

Optimal matching methods consider three different kinds of edits. First, *insertions* refer to a sequence extended by adding a state at a given position (NC-NC-DH-NC becomes NC-NC-DH-DL-NC by inserting DL at position 4). Second, *deletions* correspond to a sequence shortened by removing a specific state at a given position (NC-NC-DH-NC becomes NC-DH-NC by deleting NC at position 1 or 2). Finally, substitutions can be found where a state is changed directly, without modifying the length of the sequence (NC-NC-DH-NC becomes NC-NC-DH-DL by substituting DL for NC at position 4). Such substitutions can occur across the entire sequence. The total 'distance' between two sequences is the number of fewest insertion, deletion, and substitution edits that must be made in order to transform one sequence into the other (Table 1).

Table 1. Distances between example sequences for different insertion, deletion, and substitution costs

| **Sequence 1** | **Sequence 2** | **Distance with all costs = 1** | **Distance with in/del = 1, subst. = 5** | **Distance with in/del = 5, subst. = 1** |
|---|---|---|---|---|
| CH CH CL CH | CH CH CL CL | **1**, substitution at **position 4**: CH → CL | **2**, delete CH at 4, insert CL at 4 | **1**, substitution at **position 4**: CH → CL |
| CH CL DH CL | CL CH DL CL | **3**: sub at **3** (DH→DL), del at **1** (CH), ins at **4** (CL) | **4**, same edits, substitution penalized more | **6**, substitutions at 1, 2, 3 |
| CH CH CH CL CL | CL CL CH CH CH | **4**, delete CH at **1,2**, insert CH at **4,5** | **4**, delete CH at **1,2**, insert CH at **4,5** | **4**, substitutions at 1,2,4,5 |

Source: authors. Note: Depending on the insertion, deletion, and substitution costs, sequences may be more (or less) similar.

One important aspect of OM to note is that any location in a sequence might be shifted forward (or backward) through insertion (or deletion). This corresponds to shifting entire parts of the sequence forward (or backward) *in time*. Practically speaking, this means that places will be considered as similar if their sequences share similar *motifs*, or sub-sequence patterns of states, regardless of exactly when these *motifs* occur. As discussed by Kang et al. (2020), sequences with many common motifs may be re-aligned based on those motifs when insertion/deletion costs are low relative to substitution costs. In contrast, when insertion/deletion costs are higher, motifs become less important than the similarity of places' experiences at each point in time.

For example, consider three sequences with one motif and no conflict elsewhere. The first has a "diffuse-then-disperse" motif ("CH-DH-DL-DL") over 2008-2011, another has this same motif over 2012-2015, and a third has a different motif ("CH-DH-CL-CL") over 2008-2011. Low insertion/deletion costs prioritize the first pair's *similarity in motif*, since the same sub-sequence shows up at different time periods. High insertion-deletion costs lead OM to match the first and third sequences for their *similarity in time* – both experience "CH-DH" in 2008-09.

We focus the clustering on *motifs*, setting the in/del costs low relative to substitution costs so that places are similar when they experience similar motifs, even if the motif occurs in one place a decade before it is observed in the second place. Importantly, as Kang et al. (2020) argue, it may be unrealistic to assume that all substitutions have the same 'cost.' For instance, an edit changing CH to NC may be more 'expensive' than recoding it to CL. Thus, the similarity between states themselves can play a role in defining how expensive any single edit is. This kind of information about state similarity is often *not* considered in sequence analysis but can affect the information resulting from the clustering.

We opt to define these costs using an empirical strategy, informed by the theory of the SCDi: we set substitution costs using the observed transition frequencies in the data, thereby ensuring that substitutions are ‘cheaper’ between states that often transition into one another. This is formally given by the probability of observing state $j$ at $t+1$ when $t=i$: $2-p(i|j)-p(j|i)$ where $p(i|j)$ is the probability of transition between states $i$ and $j$. The empirical transition frequencies are shown in Figure 2.

Figure 2. Empirical transition rates for all sequences

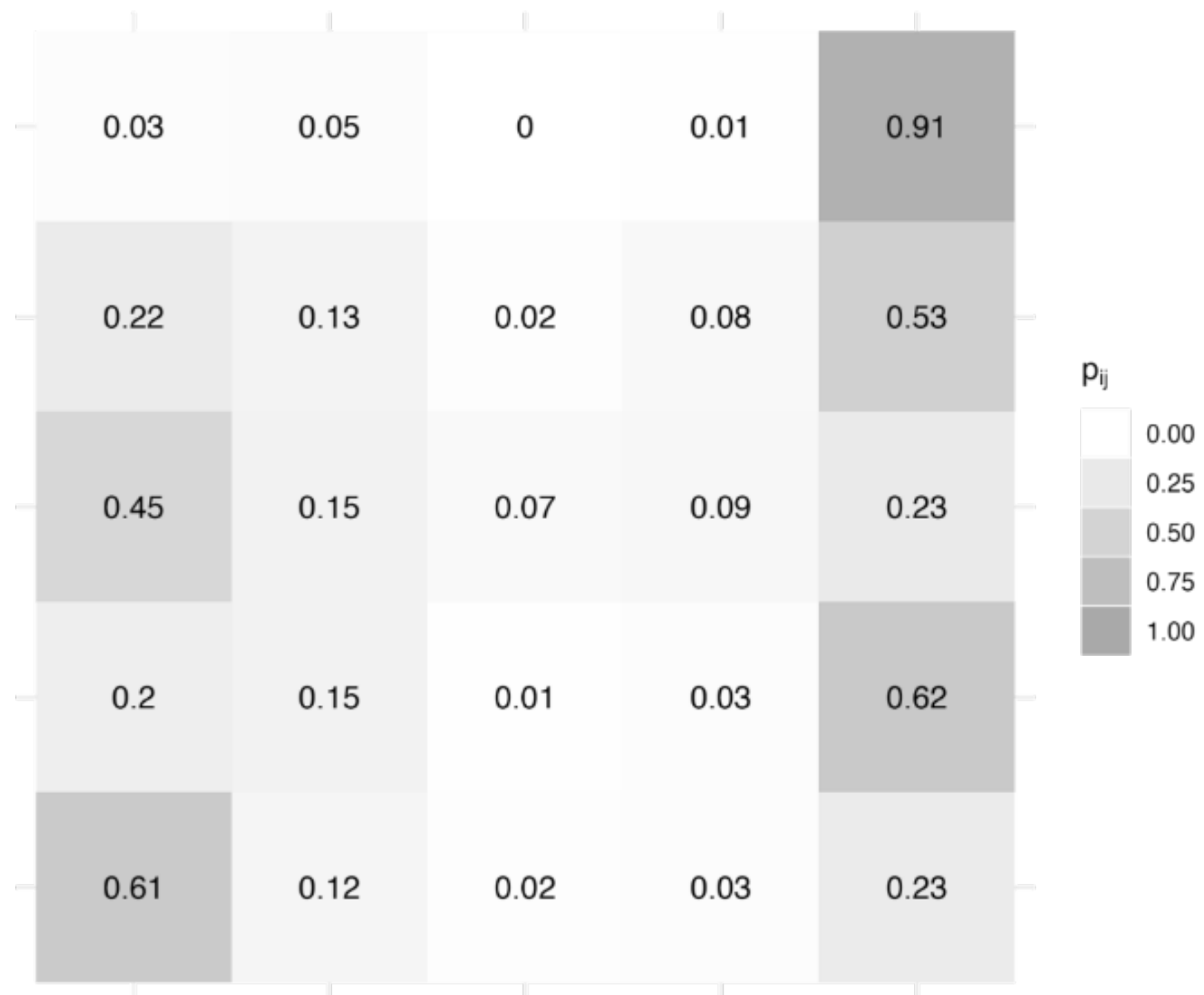

Source: authors.

*3.4. Cluster maps from OM sequence similarities*

After OM provides edit distances in the sequences for each grid cell, we use Ward’s hierarchical clustering to group similar sequences together in order to infer groups of places that experience similar motifs. Ward’s method is an agglomerative or ‘bottom-up’ clustering algorithm. Starting with each sequence as a unique cluster, sequences, and subsequently clusters of sequences, are merged at each step in such a way as to minimize the within cluster squared edit distance. We specified the number of clusters using the dendrogram, closely inspecting the results of clustering from 2 to 8 clusters. We found that the 6-cluster solution (with an additional seventh cluster implied for cells that experienced no conflict) yielded the most intuitive results. These types can be grouped in two major branches, which we call Shorter-Term (Type 1-4) and Longer-Term (Type 5-6) conflict trajectories, based on how durable conflict is once it has begun, the rates at which conflict states transition to each other or to no conflict, and the relative length of a cycle of violence (Figure 3). Their transition matrices are presented in Figure 4 and 5.

Figure 3. Trajectory branches and subtypes

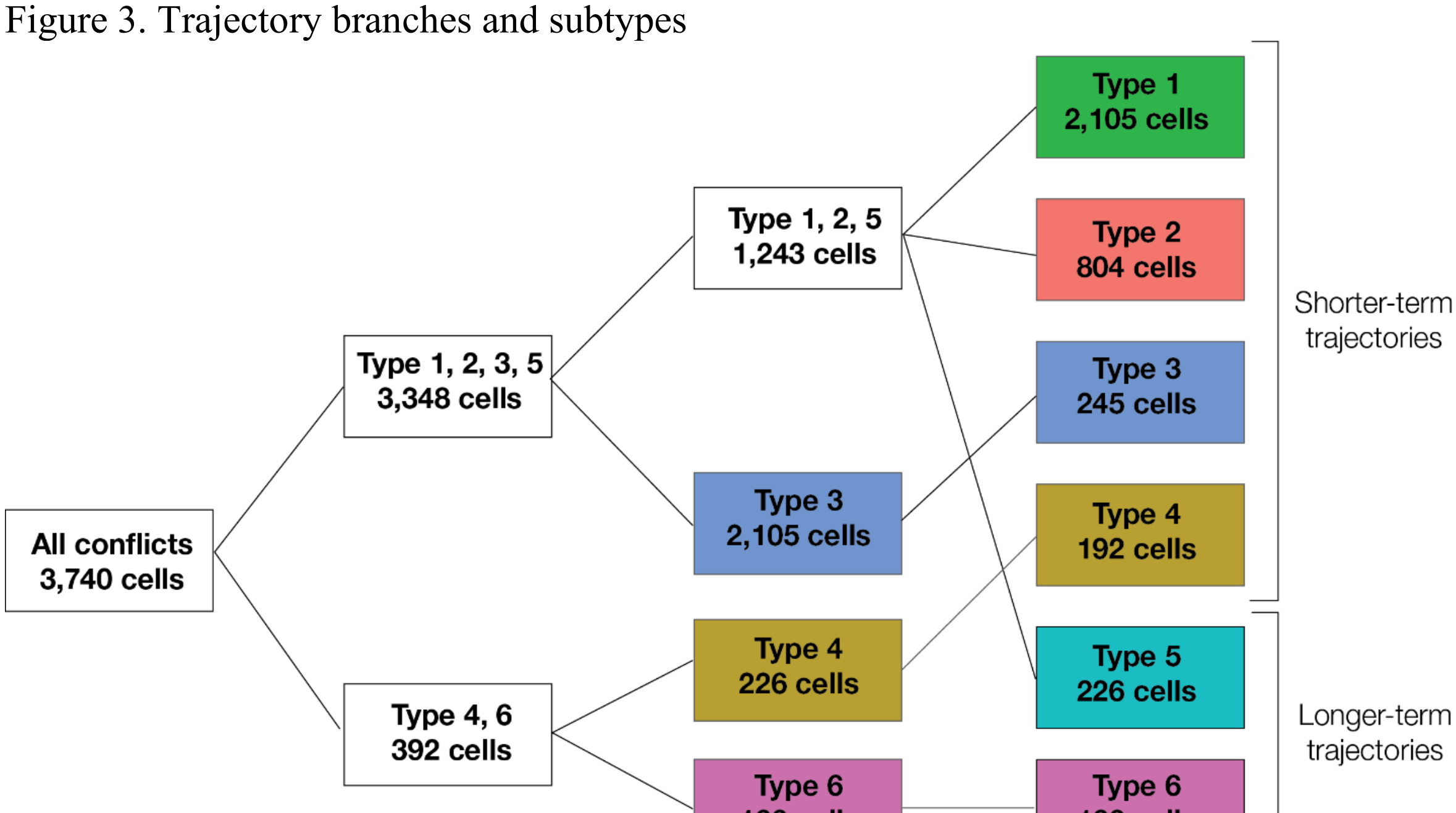

Source: authors.

Figure 4. Transition probability matrices for Shorter-Term conflict subtypes (1-4)

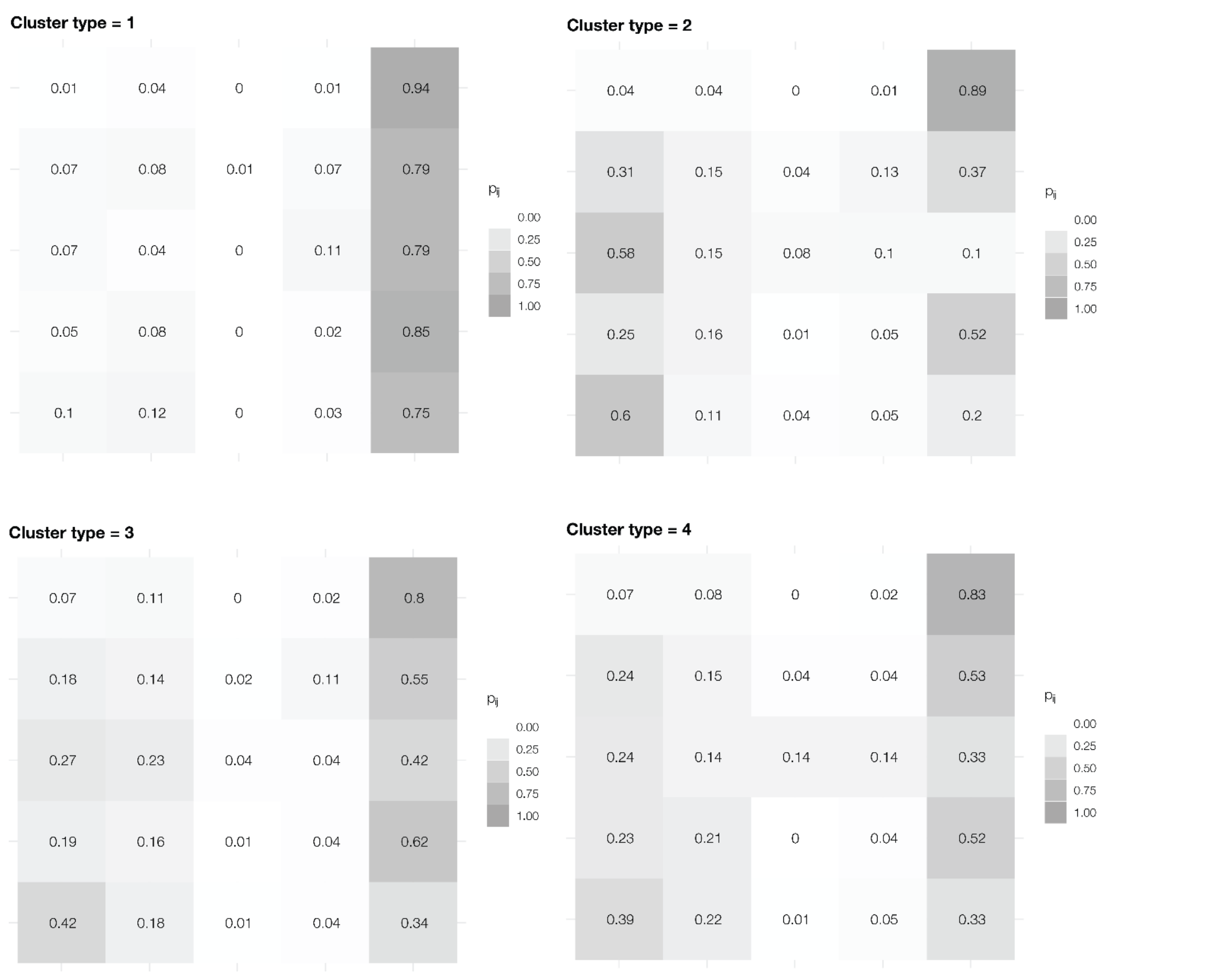

Source: authors.

Figure 5. Transition probability matrices for Longer-Term conflict subtypes (5-6)

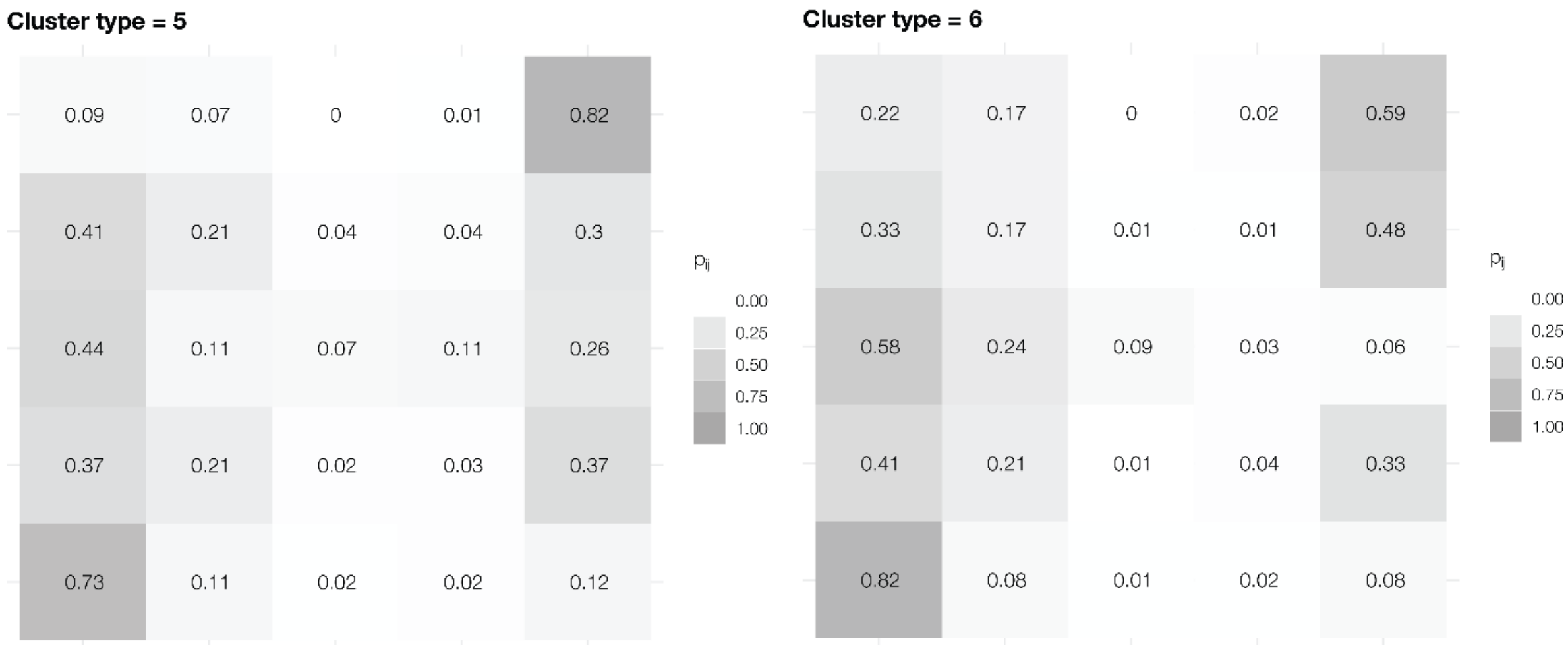

Source: authors.

From this set of clusters, we present the categorical choropleth maps showing the sequence cluster for each grid cell and calculate join count statistics to characterize the frequency of co-location between cluster types. We use the multitype join count statistical presented by Cliff and Ord (1981), which tests whether the distinct values of a multinominal categorical variable display spatial autocorrelation. Formally, the total number of joins, *jtot*, is given by:

$$Jtot = \frac{1}{2}\sum_{i}\sum_{j} w_{ij}\theta_{i,j}$$

where $w_{ij}$ are the elements of a binary spatial weights matrix that specifies the network of connectivity between locations and $\theta_{i,j} = 1$ if the states in locations $i$ and $j$ are of a different type, otherwise $\theta_{i,j} = 0$. *Jtot* can then be compared to the expected number of observed joins for each potential combination of joins which are estimated under non-free sampling from the observed frequency of the states and their variances. A z-value can then be used to indicate the extent of clustering. Together, this allows us to see (and verify) how grid cells in each cluster type locate together within the map (Table 2).

Table 2. Pair wise z-scores for multitype join count analysis

| **Subtype** | **1** | **2** | **3** | **4** | **5** | **6** |
|---|---|---|---|---|---|---|
| 1 | 19.83 | -6.52 | -5.06 | -11.47 | -4.67 | -1.59 |
| 2 | - | -1.73 | -15.53 | NA | -5.74 | -14.46 |
| 3 | - | - | 10.57 | 2.29 | 3.63 | 11.06 |
| 4 | - | - | - | 25.61 | -5.37 | -0.5 |
| 5 | - | - | - | - | 7.36 | 2.12 |
| 6 | - | - | - | - | - | 13.02 |

Source: authors.

Further, we estimate the mean violence stopping time (MVST) for each cluster using the average hitting times from any violent state to NC weighted by the fraction of violent states at the start. We conceptualize this as an *empirical* description of the typical time it took for locations to decay from a violent state into a no-conflict state within our data, and do not suggest that the distribution of violent states within clusters will stay the same into the future.

**4. Results**

Our analysis identifies clear empirical regularities in how conflicts evolve across space and time in Africa. By clustering the sequences of state transitions for each grid cell, we generate a typology of six distinct conflict trajectories. Figure 6 and Table 3 below summarize locations and key metrics for all the types. We discuss each type in turn, highlighting the defining temporal dynamics, spatial distribution, and implications for conflict persistence. To retain transparency, we report both the descriptive statistics (transition probabilities, self-transition rates, mean violence stopping times) and their broader interpretive significance.

Figure 6. Conflict trajectory subtypes across Africa, 1997-2024

Source: authors.

Table 3. Key characteristics and transition patterns of shorter-term conflict trajectories

| | **Most frequent sequence start** | **Most frequent repetition*** | **Most frequent transition^** | **Most frequent sequence terminus** | **Mean violence stopping time (years)** |
|---|---|---|---|---|---|
| *All cells (n=3,740)* | *NC>CL 5%* | *CH>CH 61%* | *DH> CH 45%* | *CL>NC 61%* | *2.94* |
| Type 1 (n=2,105) | NC>CL 4% | CH>CH 10% | CH>CL 12% | CL>NC 85% | 1.23 |
| Type 2 (n=804) | NC>CH/CL 4% | CH>CH 60% | DH>CH 58% | CL>NC 52% | 3.67 |
| Type 3 (n=245) | NC>CL 11% | CH>CH 42% | DH>CL 27% | CL>NC 62% | 2.17 |
| Type 4 (n=192) | NC>CL 8% | CH>CH 39% | DH/DL>CH 24% | DL>NC 53% | 2.38 |
| Type 5 (n=226) | NC>CH 9% | CH>CH 73% | DH>CH 44% | CL>NC 37% | 5.89 |
| Type 6 (n=166) | NC>CH 22% | CH>CH 82% | DH>CH 58% | DL>NC 48% | 8.41 |

Source: authors. Notes: * Refers to the most frequent state that transitions to itself, excluding NC to NC. ^ Refers to the most frequent state that transitions to a different type, excluding NC to NC.

### *4.1. Shorter-Term conflict trajectories*

The first branch encompasses four trajectory types that account for most grid cells experiencing conflict. These shorter-term patterns are marked by relatively rapid transitions back to a NC state, with mean violence stopping times (MVST) ranging from just over one to four years. Collectively, Types 1–4 account for roughly 80% of all conflict-affected cells in the dataset.

- **Ephemeral conflicts (type 1)** represent the most frequently observed trajectory (n = 2,105). Here, violence emerges but dissipates quickly, often within a single year. The dominant transition is from CL to NC, which occurs in 85% of observed sequences. Across the subtype, more than three-quarters of all transitions from violent states lead directly to NC, and the NC→NC self-transition rate is the highest among all groups (94%). The MVST is just 1.23 years, indicating that once violence emerges it tends to be extinguished almost immediately. Geographically, Type 1 appears widely across Africa, often on the peripheries of longer-running conflict zones. These are places where local conditions permit outbreaks but not persistence, suggesting that insurgent networks, state weakness, or political grievances are insufficient to sustain cycles of violence.

- **Clustered conflicts (type 2)** is the second most common subtype (n = 804). In these areas, violence tends to emerge in clustered forms, often around strategic sites such as towns, transport routes, or resource-rich localities. The CH→CH self-transition rate is extremely high (60%), indicating repeated cycles of intense violence. The most common path to de-escalation occurs via CL→NC transitions (52%), although this pathway is less frequent than in Type 1. MVST averages 3.67 years, the third highest among all subtypes, suggesting that once violence emerges it remains a recurrent feature. Type 2 is spatially concentrated in active theaters of armed conflict, forming belts across the Sahel, eastern DRC, northern Mozambique, and parts of Libya. These areas are typically marked by persistent insurgencies or recurring battles for territorial control.

- **Past conflicts (type 3)**, with n = 245, captures locations with histories of armed conflict but little contemporary recurrence. The most common starting sequence is NC→CL (11%), reflecting brief flare-ups after long peaceful intervals. The most frequent transition is CL→NC (62%), which dominates the trajectory. MVST is 2.17 years, reflecting short cycles when violence does occur. Type 3 is most prominent in countries such as Liberia, Sierra Leone, Zimbabwe, and Algeria, contexts where large-scale conflicts have ended and not reignited. These patterns suggest a durable peace, though one occasionally punctuated by small-scale violent episodes.

- **Unstable conflicts (type 4)** are the least common among the Shorter-Term branch (n = 192) but has the most complex transition dynamics. Here, DH states act as intermediaries: DH frequently transitions to CH (24%) or DL (27%), creating oscillations among states before eventual resolution. The DH category is much more active in this subtype than in others, serving as a pivot rather than a terminal state. MVST averages 2.38 years, but the volatility of transitions suggests conflicts are not firmly contained. Spatially, Type 4 often co-occurs with Type 2 in regions of ongoing instability, such as the Sahel, eastern DRC, and parts of Nigeria.

Taken together, the Shorter-Term branch illustrates how most conflict in Africa manifests as localized, short-lived episodes. However, the persistence of Types 2 and 4 shows that clustered or unstable cycles can create pockets of recurrent instability. These types highlight how certain geographies, especially those already engaged in broader conflicts—serve as sites of repeated violence even within otherwise short-term patterns.

*4.2. Longer-Term conflict trajectories*

In contrast, the Longer-Term branch captures the subset of African conflict zones where violence becomes entrenched and highly resistant to resolution. These types are characterized by the dominance of CH states, which persist over time and show very low probabilities of transitioning to NC.

- **Enduring conflicts (type 5)** are characterized by exceptionally durable cycles of violence (n = 226). CH states dominate, with CH→CH self-transitions at 73%—among the highest of all subtypes. By comparison, transitions to NC are rare (12%), and CL→NC transitions (37%) are much less common than in the shorter-term types. MVST is 5.88 years, indicating that once violence emerges, it is likely to persist for half a decade or more. Geographically, Type 5 aligns closely with the most enduring African conflict theaters: Mali and Burkina Faso, the Lake Chad Basin, Darfur, Somalia, and the eastern DRC. These are zones where insurgent organizations are deeply entrenched, grievances remain unresolved, and state capacity is weak or contested.

- **Balanced conflicts (type 6)** are the rarest (n = 166) but most persistent conflict trajectory. It is defined by the overwhelming dominance of CH states: 82% of CH cells remain in CH year to year, and only 8% transition to NC. Violence often begins with NC→CH transitions (22%), the highest such rate across all types, indicating abrupt and direct escalation into high-intensity clustered states. MVST is 8.41 years, making these the longest cycles observed. Balanced conflicts typically occur where belligerents are evenly matched, producing prolonged confrontations with no decisive resolution. The eastern Great Lakes region exemplifies this subtype, where states and militias backed by neighboring governments have fought recurrent battles without establishing durable peace. Similar patterns are evident in Nigeria, Darfur, Somalia, and Egypt.

The Longer-Term branch highlights the path dependency of clustered high-intensity violence. Once a locality enters CH states, de-escalation is extremely unlikely. These areas represent the hardest conflicts to resolve and remain the central challenges for peacebuilding efforts across Africa.

*4.3. Spatial clustering of trajectories*

Beyond typology, spatial analysis reveals significant clustering of conflict types (Figure 7). Using join count statistics, we find that similar trajectories are geographically contiguous more often than expected by chance. Shorter-term types (especially Types 1 and 3) are scattered across nearly all African countries, often surrounding pockets of more persistent violence. Longer-term types (Types 5 and 6) dominate in transborder hotspots, including the central Sahel, the Lake Chad Basin, and the eastern Great Lakes.

This spatial pattern demonstrates that conflict trajectories are not randomly distributed but shaped by neighborhood effects. Borderlands emerge as key linking spaces, where enduring conflict types extend across national boundaries and overlap with unstable shorter-term patterns. These findings align with research showing the importance of border regions as recruitment bases, logistical hubs, and safe havens (Arsenault & Bacon, 2015; Dowd, 2017; Radil et al., 2022). Our results confirm empirically that such areas are not only sites of conflict onset but also enduring nodes in transregional conflict complexes.

Figure 7. Violent trajectories in Africa, by type, 1997-2024

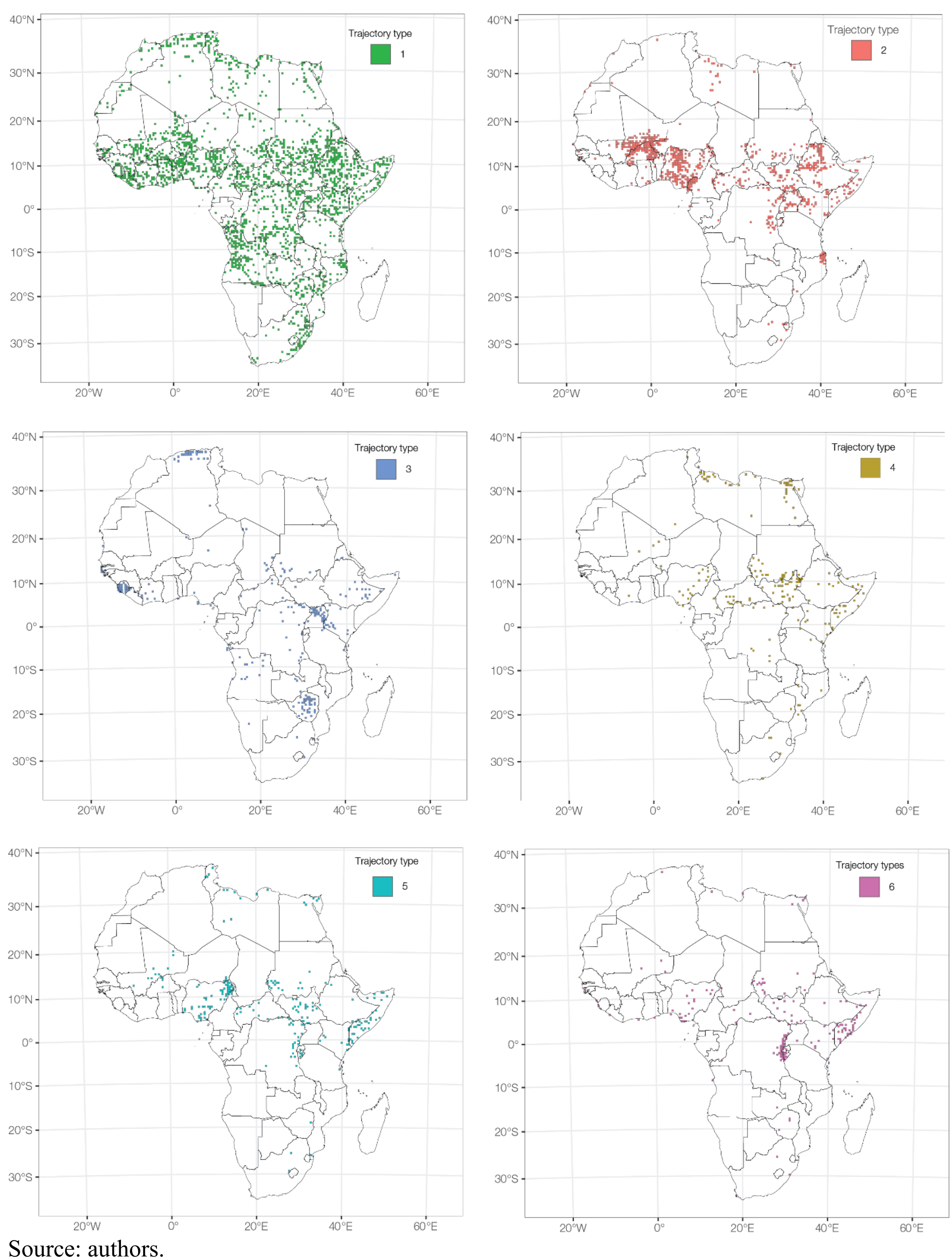

Source: authors.

## 5. Discussion

This study provides new evidence about the long-term spatial and temporal dynamics of armed conflict in Africa. By applying sequence analysis to more than two decades of event data, we identify six recurrent conflict trajectories and demonstrate that these trajectories are not randomly distributed but instead cluster geographically, often around borders and in historically unstable regions. Most conflicts are short-lived, returning to "no conflict" within one to three years. Yet once violence becomes clustered and high intensity, it is likely to persist for extended periods, often exceeding half a decade. Spatial analysis reveals that these trajectories are geographically patterned, with shorter-term conflicts widely distributed but longer-term trajectories concentrated in borderland regions.

These findings demonstrate that conflict persistence is strongly path dependent: once a locality enters a high-intensity clustered state, it is unlikely to de-escalate quickly. They also show that the trajectories of neighboring localities are systematically associated, producing regional complexes of violence that cut across state boundaries. Together, these results suggest that conflict in Africa cannot be fully understood through discrete transitions alone but requires attention to the broader life cycles of violence within and between places.

### *5.1. Rethinking conflict trajectories and persistence*

One of the central findings of this study is that conflicts tend to follow identifiable temporal and spatial trajectories rather than simply oscillating between violent and non-violent states. This challenges the implicit assumptions of many regression-based or transition-oriented approaches, which model conflict as discrete events without reference to their cumulative sequencing. Our results show that this simplification risks obscuring important differences between short-lived outbreaks and enduring conflict zones.

In particular, the persistence of CH states across Types 5 and 6 underscores the path dependency of violence. Once a locality enters such a state, it is highly unlikely to de-escalate quickly. The MVST for Type 6 exceeds eight years, highlighting that certain forms of violence become self-reinforcing and deeply embedded in local dynamics. These findings complement and extend prior research on conflict duration (Buhaug et al., 2009; Toft & Zhukov, 2015), which has emphasized structural factors such as terrain, resource endowments, or rebel capacity. Our approach shows that persistence is also a function of the sequence of prior states: high-intensity clustered violence is uniquely resistant to change, even when other contextual factors are similar.

By contrast, the Shorter-Term trajectories (Types 1–4) reveal that most violent episodes in Africa remain ephemeral or transitional. Type 1 cells, for example, return to "no conflict" within little more than a year, suggesting that sporadic outbreaks often reflect opportunistic actions or local disturbances rather than systemic instability. Types 2 and 4, however, demonstrate that clustered or unstable cycles can generate recurrent instability even when violence is eventually suppressed. These findings show that conflict persistence cannot be understood solely by examining whether violence occurs in a given year; it is the temporal ordering of states that determines whether violence is fleeting or enduring.

*5.2. Diffusion, neighborhood effects, and borderlands*

A second key contribution of our analysis is to demonstrate that trajectories are spatially clustered and that neighboring localities frequently share similar conflict sequences. This supports longstanding arguments that conflict is contagious and diffuses across space (O'Loughlin & Raleigh, 2008; Schutte & Weidmann, 2011), but it adds new nuance by showing that diffusion operates not just through isolated transitions but through the alignment of broader trajectories.

Borderlands emerge as especially significant in this respect. Our results identify clusters of enduring and balanced conflicts that span multiple states, including the Sahel, Lake Chad, and the Great Lakes. These zones align with prior accounts of borderlands as logistical hubs, recruitment pools, and safe havens (Arsenault & Bacon, 2015; Dowd, 2017, Radil et al., 2022; OECD, 2023). Yet our analysis goes further in showing how border regions serve as long-term conduits of trajectory types. For instance, Type 5 enduring conflicts are heavily concentrated in Mali and Burkina Faso but extend seamlessly across national boundaries, while Type 6 balanced conflicts define much of the eastern DRC and adjacent territories. These findings illustrate that neighborhood effects are not merely short-term contagion but reflect the potential long-run comingling of some conflict trajectories across space.

This spatial clustering has theoretical implications for understanding regional conflict complexes (Buhaug & Gleditsch, 2008; Braithwaite & Jeong, 2017). It suggests that interventions in one locality are unlikely to succeed without considering the broader trajectory patterns of neighboring areas. A locality's conflict cycle cannot be disentangled from those of its neighbors when shared borders and transnational networks create enduring interdependencies.

*5.3. Contributions to methodological debates*

The application of sequence analysis represents a methodological innovation in conflict research. While sequence methods are increasingly used in other branches of geography and the social sciences, they remain rare in the study of political violence. Our study demonstrates that sequence analysis can capture both the timing and ordering of violent events, producing a typology of trajectories that can complement and extend traditional event-history or diffusion models.

A key advantage of this approach is its ability to differentiate between short-lived irregularities and sustained shifts in conflict dynamics. Transition-based methods treat each move between violent and non-violent states as equivalent, regardless of whether it occurs in a stable or unstable sequence. By contrast, sequence analysis situates each transition within the broader history of a locality, enabling us to distinguish between "noise" and systemic change. This has important implications for prediction: rather than relying only on covariates or past event counts, analysts can incorporate information about where a locality sits in its longer conflict trajectory.

Another strength is the capacity to compare trajectories across locations. By clustering sequences, we identify common motifs and group localities into trajectory types. This provides a framework for generalization that is often missing in highly localized studies. At the same time, it avoids the homogenizing assumptions of global regression models, instead acknowledging that conflicts evolve differently in different contexts.

### *5.4. Policy implications*

Our findings also have implications for policy and practice. While it is common to call for "tailored interventions," our results demonstrate concretely why such tailoring is necessary. Distinct conflict trajectories imply different intervention logics.

Ephemeral conflicts (Type 1) often resolve themselves quickly; heavy-handed interventions in these contexts may be unnecessary or even counterproductive. Instead, policies could focus on preventing opportunistic flare-ups from escalating into more persistent cycles. Clustered or unstable conflicts (Types 2 and 4) suggest the need for repeated stabilization efforts in specific localities. These conflicts often revolve around key towns, transport hubs, or resource sites. Interventions here should target the local political economy of violence and provide mechanisms for repeated conflict management.

Enduring and balanced conflicts (Types 5 and 6) pose the greatest challenge. The persistence of CH states implies that once violence escalates to this level, de-escalation is unlikely without sustained, coordinated intervention. These areas often cut across borders, making unilateral strategies ineffective. Regional cooperation, integrated peacebuilding, and long-term governance reforms are essential.

In each case, recognizing whether a locality is on an ephemeral, unstable, or enduring trajectory matters for resource allocation and strategic planning. Early warning systems, such as ViEWS (Hegre et al., 2019), could incorporate trajectory typologies to improve predictions about which areas are at greatest risk of sustained violence. Likewise, peacebuilding initiatives could prioritize regions showing signs of escalation into CH states, treating them as critical intervention points.

### *5.5. Limitations and future research*

Several limitations of our analysis should be noted. First, while ACLED provides extensive coverage of conflict events, it is not free from reporting biases (Eck, 2012; Miller et al., 2022). These biases may influence the precise categorization of states or the length of observed trajectories. Replicating our analysis with alternative disaggregated event datasets would be a useful next step.

Second, our focus has been on the patterns of trajectories rather than their causal determinants. While we show that certain trajectories are more common in certain regions, we do not directly test which factors explain why a locality falls into one trajectory type rather than another. Future research could integrate sequence analysis with explanatory models that incorporate governance quality, economic structures, or social networks.

Third, while our approach captures spatial clustering, it does not specify the mechanisms of diffusion. Do conflicts spread through refugee flows, arms trafficking, or insurgent mobility? Are trajectories synchronized because of cross-border ethnic networks or because state weakness is regionally correlated? Disentangling these mechanisms requires mixed-method approaches that combine sequence analysis with qualitative or network-based investigations.

Finally, while our study focuses on Africa, the approach could be extended to other regions. Applying sequence analysis to conflicts in the Middle East, South Asia, or Latin America would reveal whether similar trajectory types emerge elsewhere, or whether African conflicts are distinctive in their sequencing.

## 6. Conclusion

This study applied spatial sequence analysis to two decades of political violence in Africa, identifying six recurrent conflict trajectories that differ in duration, intensity, and spatial distribution. The findings demonstrate that most conflicts are short-lived, returning quickly to non-violent states, but that once violence escalates into clustered high-intensity forms, it is highly path dependent and resistant to resolution. A second key contribution is the recognition of spatial clustering. Conflict trajectories align across neighboring localities, particularly in border regions, creating enduring regional complexes of violence. These patterns underscore the importance of analyzing conflict not only as discrete events but as evolving life cycles embedded in wider geographic contexts.

Methodologically, the study highlights the utility of sequence analysis for capturing the temporal rhythms and spatial interdependencies of violence. Practically, it suggests that interventions should be tailored to conflict trajectories: monitoring for ephemeral outbreaks, stabilization in unstable zones, and coordinated long-term strategies for enduring conflicts. By shifting attention from transitions to trajectories, this work provides a framework for understanding the persistence and diffusion of violence, offering both theoretical insights and actionable implications for peacebuilding.

Taken together, these findings and limitations suggest a research agenda that integrates temporal sequencing with spatial interdependence. Future studies might examine critical junctures of escalation: Under what conditions do ephemeral or clustered conflicts escalate into enduring trajectories? They may also examine cross-border linkages, i.e., how transnational networks shape the synchronization of trajectories across regions. Another avenue of research is to study why some localities sustain no conflict despite proximity to enduring conflicts, while others succumb to recurrent cycles. Finally, future research could examine how do peacekeeping operations, regime changes, or international interventions alter trajectory types, and are some trajectories more resistant than others?

By framing conflicts as sequences rather than isolated events, scholars can begin to answer these questions in new ways. Sequence analysis is not a replacement for other methods but a complement, offering a lens that foregrounds the temporal and spatial rhythms of violence.

**References**

Arsenault, E. G., & Bacon, T. (2015). Disaggregating and defeating terrorist safe havens. *Studies in Conflict & Terrorism*, *38*(2), 85-112.

Bormann, N. C., & Hammond, J. (2016). A slippery slope: The domestic diffusion of ethnic civil war. *International Studies Quarterly*, *60*(4), 587-598.

Braithwaite, A., & Jeong, S. (2017). Diffusion in international politics. In *Oxford Research Encyclopedia of Politics*.

Braithwaite, A., & Li, Q. (2019). Transnational terrorism hot spots: Identification and impact evaluation. In *Transnational Terrorism* (pp. 65-80). Routledge.

Buhaug, H., Gates, S., & Lujala, P. (2009). Geography, rebel capability, and the duration of civil conflict. *Journal of Conflict Resolution*, 53(4), 544-569.

Cliff, A. D., & Ord, J. K. (1981). *Spatial processes: models & applications*. London: Pion.

D'Amato, S. (2018). Terrorists going transnational: rethinking the role of states in the case of AQIM and Boko Haram. *Critical Studies on Terrorism*, 11(1), 151-172.

Delmelle, E. C. (2016). Mapping the DNA of urban neighborhoods: Clustering longitudinal sequences of neighborhood socioeconomic change. *Annals of the American Association of Geographers*, 106(1), 36-56.

Demarest, L. and Langer, A. (2022). How events enter (or not) data sets: The pitfalls and guidelines of using newspapers in the study of conflict. *Sociological Methods & Research*, 51(2), 632-666.

Dowd, C. (2017). Nigeria's Boko Haram: local, national and transnational dynamics. In *African Border Disorders* (pp. 115-135). Routledge.

Eck, K. (2012). In data we trust? A comparison of UCDP GED and ACLED conflict events datasets. *Cooperation and Conflict*, 47(1), 124-141.

Giraudy, A., Moncada, E., & Snyder, R. (2019). Empirical and theoretical frontiers of subnational research in comparative politics. *Inside Countries: Subnational Research in Comparative Politics*, 353-367.

Hegre, H., Allansson, M., Basedau, M., Colaresi, M., Croicu, M., Fjelde, H., ... & Vestby, J. (2019). ViEWS: A political violence early-warning system. *Journal of Peace Research*, 56(2), 155-174.

Kang, W., Rey, S., Wolf, L., Knaap, E., & Han, S. (2020). Sensitivity of sequence methods in the study of neighborhood change in the United States. *Computers, Environment and Urban Systems*, 81, 101480.

Levin, N., Ali, S., & Crandall, D. (2018). Utilizing remote sensing and big data to quantify conflict intensity: The Arab Spring as a case study. *Applied Geography*, 94, 1-17.

Losacker, S., & Kuebart, A. (2024). Introducing sequence analysis to economic geography. *Progress in Economic Geography*, 2(1), 100012.

Marineau, J., Pascoe, H., Braithwaite, A., Findley, M., & Young, J. (2020). The local geography of transnational terrorism. *Conflict Management and Peace Science*, 37(3), 350-381.

Metternich, N. W., Minhas, S., & Ward, M. D. (2017). Firewall? Or wall on fire? A unified framework of conflict contagion and the role of ethnic exclusion. *Journal of Conflict Resolution*, 61(6), 1151-1173.

Miller, E., Kishi, R., Raleigh, C., & Dowd, C. (2022). An agenda for addressing bias in conflict data. *Scientific Data*, 9(1), 593.

OECD (2020). *The Geography of Conflict in North and West Africa*. OECD Publishing.

OECD (2023). *Borders and Conflict in North and West Africa*. OECD Publishing.

O'Loughlin, J., & Witmer, F. D. (2012). The diffusion of violence in the North Caucasus of Russia, 1999–2010. *Environment and Planning A*, 44(10), 2379-2396.

O'Loughlin, J., & Raleigh, C. (2008). Spatial analysis of civil war violence. *The SAGE Handbook of Political Geography*, 493-508. SAGE.

Radil, S. M., Irmischer, I., & Walther, O. J. (2022). Contextualizing the relationship between borderlands and political violence: A dynamic space-time analysis in North and West Africa. *Journal of Borderlands Studies*, 37(2), 253-271.

Raleigh, C., & Linke, A. (2018). Subnational governance and conflict: An introduction to a special issue on governance and conflict. *Political Geography*, 63, 88-93.

Raleigh, C., Linke, R., Hegre, H., & Karlsen, J. (2010). Introducing ACLED: An armed conflict location and event dataset. *Journal of Peace Research*, 47(5), 651-660.

Raleigh, C., & Hegre, H. (2009). Population size, concentration, and civil war. A geographically disaggregated analysis. *Political Geography*, 28(4), 224-238.

Salehyan, I., Hendrix, C. S., Hamner, J., Case, C., Linebarger, C., Stull, E., & Williams, J. (2012). Social conflict in Africa: A new database. *International Interactions*, 38(4), 503-511.

Schutte, S., & Weidmann, N. B. (2011). Diffusion patterns of violence in civil wars. *Political Geography*, 30(3), 143-152.

Stehle, S., & Peuquet, D. J. (2015). Analyzing spatio-temporal patterns and their evolution via sequence alignment. *Spatial Cognition & Computation*, 15(2), 68-85.

Toft, M. D., & Zhukov, Y. M. (2015). Islamists and nationalists: Rebel motivation and counterinsurgency in Russia's North Caucasus. *American Political Science Review*, 109(2), 222-238.

Sundberg, R., & Melander, E. (2013). Introducing the UCDP georeferenced event dataset. *Journal of Peace Research*, 50(4), 523-532.

Vogiazides, L., & Mondani, H. (2023). Neighbourhood trajectories in Stockholm: Investigating the role of mobility and in situ change. *Applied Geography*, 150, 102823.

Walther, O. J., Radil, S. M., Russell, D. G., & Trémolières, M. (2023). Introducing the Spatial Conflict Dynamics indicator of political violence. *Terrorism and Political Violence*, 35(3), 533-552.

Walther, O. J., Radil, S. M., & Russell, D. G. (2025). The spatial conflict life cycle in Africa. *Annals of the American Association of Geographers*, 115(2), 456-477.

Zhukov, Y. M. (2012). Roads and the diffusion of insurgent violence: The logistics of conflict in Russia's North Caucasus. *Political Geography*, 31(3), 144-156.